\documentclass[runningheads]{llncs}
\usepackage{graphicx}
\usepackage{epstopdf}
\usepackage{cite}
\usepackage[cmex10]{amsmath}
\usepackage{amssymb}
\usepackage{array}
\usepackage{float}
\usepackage{textpos}
\usepackage{subfigure}
\usepackage{enumerate}
\usepackage{lscape}
\usepackage{url}
\usepackage{color}
\usepackage{xfrac}
\usepackage{cases}
\usepackage{upgreek}
\usepackage{bm}

\usepackage{booktabs}

\usepackage{tabu}

\usepackage{tabularx}
\usepackage{threeparttable} 
\usepackage{multirow} 
\usepackage{rotating} 
\usepackage{bigstrut} 
\usepackage{rotating} 
\usepackage{hhline}
\newcolumntype{b}{>{\hsize=.25\hsize}X}
\newcolumntype{h}{>{\hsize=.5\hsize}X}
\newcolumntype{m}{>{\hsize=.75\hsize}X}

\usepackage{algorithm}
\usepackage{algpseudocode}
\usepackage{pifont}

\usepackage{hyperref}

\usepackage{cleveref}

\begin{document}
\title{Cybersecurity Assessment of the Polar Bluetooth~Low~Energy Heart-rate Sensor}
\titlerunning{Cybersecurity Assessment of the Polar BLE Heart-rate Sensor}
%
\author{S. Soderi\inst{1}}
\authorrunning{S. Soderi}
%
\institute{IEEE Senior Member \\
\email{soderi@ieee.com}}
\maketitle              
%
\begin{abstract}
Wireless communications among wearable and implantable devices implement the information exchange around the human body. Wireless body area network (WBAN) technology enables non-invasive applications in our daily lives. Wireless connected devices improve the quality of many services, and they make procedures easier. On the other hand, they open up large attack surfaces and introduces potential security vulnerabilities. Bluetooth low energy (BLE) is a low-power protocol widely used in wireless personal area networks (WPANs).
This paper analyzes the security vulnerabilities of a BLE heart-rate sensor. By observing the received signal strength indicator (RSSI) variations, it is possible to detect anomalies in the BLE connection. The case-study shows that an attacker can easily intercept and manipulate the data transmitted between the mobile app and the BLE device. With this research, the author would raise awareness about the security of the heart-rate information that we can receive from our wireless body sensors.
\end{abstract}

\newpage
\noindent\rule{8.4cm}{1pt}\\
Please cite this version of the paper:\\

Soderi, S. Cybersecurity assessment of the polar bluetooth low energy heart-rate sensor. In \textit{Body Area Networks: Smart IoT and Big Data for Intelligent Health Management: 14th EAI International Conference, BODYNETS 2019}, Florence, Italy, October 2-3, 2019, Proceedings 14 (pp. 252-265). Springer International Publishing. https://doi.org/10.1007/978-3-030-34833-5\_20
\\

You may use the following bibtex entry:
\begin{verbatim}
@InProceedings{Soderi2019BLEattack,
author="Soderi, S.",
editor="Mucchi, Lorenzo
and H{\"a}m{\"a}l{\"a}inen, Matti
and Jayousi, Sara
and Morosi, Simone",
title="Cybersecurity Assessment of the Polar Bluetooth 
Low Energy Heart-Rate Sensor",
booktitle="Body Area Networks:  Smart IoT and Big Data 
for Intelligent Health Management",
year="2019",
publisher="Springer International Publishing",
address="Cham",
pages="252--265",
isbn="978-3-030-34833-5",
doi=10.1007/978-3-030-34833-5_20
}

\end{verbatim}
\noindent\rule{8.4cm}{1pt}

\section{Introduction}
\label{SEC:Introduction}
In the last two decades, Bluetooth became very popular in the short-range communications. Every smartphone, tablet and personal computer embeds this technology. At the same time, many wireless sensors such as fitness sensors, smartwatches, headsets, wireless medical devices (WMDs) rely on Bluetooth to exchange data with the user's smartphone or tablet. Today, wireless body area networks (WBANs) collect humans' information through Bluetooth low energy (BLE) sensors nodes. BLE is thus becoming de-facto a key wireless technology and users leave that interface always enabled on their devices. BLE was introduced as \emph{Wibree} by Nokia in $2006$~\cite{2018:active-mitm-tal}. Today, BLE is the dominant technology to convey efficiently data in body networks using coin cell battery-powered devices.

Though the advantages offered by any WBAN are substantial, it makes one of the prime targets for security threats and more particularly for users' privacy.  WBANs are commonly used to track health and fitness data and it raises the interest of cyber-criminals to this kind of information.

Fitness wearable devices collect human`s information, and they are designed to be worn all day. The user reads these data through his smartphone, tablet or even by his smartwatch. Historically,  these fitness trackers have numerous security vulnerabilities and the wireless sensor or even the software application may disclose users' data. It is clear that it might have important privacy implications~\cite{2014:Fitbit-security,2016:BLE-privacy-issue}. 
In the literature, there are several contributions that deals with security aspects in WBAN health-care applications. Indeed, the security weaknesses of WMDs can lead to a high risk for the patient's safety~\cite{2013:SecThreatsMedical}. On the other hand,  standardization bodies have already adopted security solutions. The IEEE802.15.6 defines different levels of security throughout the encryption and authentication of the data~\cite{2012:IEEE802156}. Moreover,  the European Telecommunications Standard Institute (ETSI) takes into account the security in the smart body area networks (SmartBANs)~\cite{2018:etsi_tr_smartban}. SmartBANs are used for the collection, processing, and transmission of patient's data. It is crucial that the information must be securely treated~\cite{2018:mucchi_security_healthdevices}.
The rapid proliferation of wireless implantable medical devices (WIMDs) coupled with their increasing features is raising the risk for patients~\cite{2018:security-implantable-med-dev}. 

The utilization of wireless technology  makes the data prone to being eavesdropped, modified and injected. This increases concerns about the privacy of the information managed in WBANs. In this paper, the author is considering man-in-the-middle (MitM) attack in a BLE WBAN fitness scenario. Observing the received signal strength indicator (RSSI) variations, this study proposes a mechanism to detect MitM attacks.

The rest of this paper is organized as follows. Section~\ref{SEC:BLE} overviews the BLE specifications. Section~\ref{SEC:Fitness-scenario} describes the security vulnerabilities of a BLE heart-rate sensor and the results of a MitM attack. Then, the paper proposes security countermeasures. Finally, conclusions are presented in Section~\ref{SEC:conclusions}.

\section{Bluetooth Low Energy (BLE)}
\label{SEC:BLE}

\subsection{BLE Core Specifications}
Bluetooth is an open standard used for short-range communications. This wireless technology operates in the $2.4$~GHz ISM band and it is primarily used for consumer, medical and personal devices~\cite{bluetooth-radio}. With Bluetooth users can create personal ad-hoc networks to transfer any kind of data. Above $5$ billion Bluetooth devices are expected to be shipped within $2022$~\cite{bluetooth-market-2018}. At the time of writing, Bluetooth $5.0$ is the most recent version and it is rapidly adopted in smartphones. Despite the Bluetooth Special Interest Group (SIG) releases newer versions, even older ones are currently in use and is common to find Bluetooth $4.1$ and $4.2$ in commercial devices~\cite{bluetooth-sig,bluetooth-market-2018}. The Bluetooth architecture specifies two forms: basic rate/enhanced data rate (BR/EDR) and low energy (LE)~\cite{bluetooth-core-specs}. This paper refers to the BLE standard.

Table~\ref{TAB:Bluetooth-radio} shows a comparison of the lower layers between BLE and Bluetooth BR/EDR. This comparison indicates a different usage of the radio spectrum. BLE splits spectrum into $40$ channels: $3$ advertising channels to establish connections and $37$ channels to transmit data. Furthermore, BLE devices are designed to send short bursts of data rather than a continuous data stream. It makes BLE ideal for sensor applications. BLE devices consume very low energy in comparison to other wireless technologies. 
 
%
\begin{table}[!t]
	\centering
	\begin{threeparttable}
		\renewcommand{\arraystretch}{1.2}
		\caption{Lower layers comparison between BLE and Bluetooth BR/EDR~\cite{bluetooth-radio}}
		\label{TAB:Bluetooth-radio}
		\begin{tabular}{|c|c|c|}
			\hline
			\bfseries Characteristic & \bfseries Bluetooth LE (BLE) & \bfseries Bluetooth BR/EDR \\
			\hline
			Frequency band  &  2.4 GHz & 2.4 GHz \\
			\hline
			Channels   & \begin{tabular}[c]{@{}c@{}} 40 channels with 2 MHz spacing \\ (3 advertising ch./37 data ch.)\tnote{1}\end{tabular} & 79 channels with 1 MHz spacing \\
			\hline
			Channel usage  & FHSS & FHSS  \\
			\hline
			Modulation   & GFSK & GFSK, $\frac{\pi}{4}$ DQPSK, 8DPSK  \\
			\hline
			Max data-rate   &  2 Mbps & 3 Mbps \\
			\hline
			Max Tx power  & 100 mW & 100mW  \\
			\hline
			Power consumption &  (0.01 $\div$ 0.05)$\cdot $(1)\tnote{2}     &	(1)\tnote{2}	\\
			\hline
			Network topologies      & \begin{tabular}[c]{@{}c@{}} Point-to-Point\tnote{3}, \\  Broadcast, Mesh\end{tabular} & Point-to-Point\tnote{3} \\
			\hline
			Connection      & Short burst data transmission & Continuous data stream \\
			\hline
			Typical range   & 30 m & 50 m \\
			\hline
		\end{tabular}
		\begin{tablenotes}
			\item [1] Advertising channels: ch. 37 (2402 MHz), ch. 38 (2426 MHz) and ch. 39 (2480 MHz);
			\item [2] (1) is the reference value;
			\item [3] Including piconet.
		\end{tablenotes}
	\end{threeparttable}
\end{table}
%
\begin{figure}[!h]
	\centering
	\includegraphics[width=0.75\textwidth]{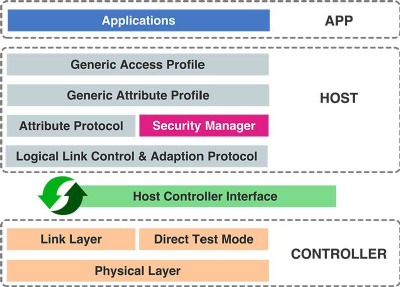}
	\caption{Architecture of BLE.} \label{FIG:BLE-STACK}
\end{figure}

BLE is a full protocol stack. It is a combination of hardware parts and software layers~\cite{bluetooth-core-specs}. As shown in Figure~\ref{FIG:BLE-STACK},  BLE architecture is organized in three major blocks: applications, host and controller. The user \emph{application} defines the interface with the Bluetooth stack. \emph{Host} block consists of the upper layers, whereas \emph{controller} includes lower layers. Host and controller communicate through the host controller interface (HCI). This division makes possible to interface many hosts with a single controller by using the HCI.

The generic access profile (GAP) layer controls the role and connection of a BLE device. BLE  specifications define GAP roles as follows~\cite{bluetooth-core-specs}
\begin{itemize}
	\item \textbf{Broadcaster}: a device that only sends advertising events;
	\item \textbf{Observer}: a device that only receives advertising events;
	\item \textbf{Peripheral}: a device that accepts the establishment of a LE physical link using the connection establishment procedure;
	\item \textbf{Central}: a device that initiates the establishment of a physical connection.
\end{itemize}
%
\begin{figure}[!b]
	\centering
	\includegraphics[width=0.75\textwidth]{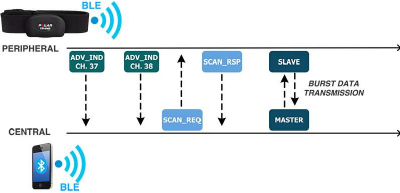}
	\caption{BLE connection flow.} \label{FIG:BLE-CONN}
\end{figure} 

The logical link control and adaptation protocol (L2CAP) layer plays a central role in Bluetooth stack.  It takes multiple protocols from the upper layers and encapsulates them into the standard BLE packet format and vice-versa. L2CAP layer is in charge or routing two main protocols: the attribute protocol (ATTP) and the security manager (SM).

Figure~\ref{FIG:BLE-CONN} shows the connections flow between master and slave  BLE devices. In a fitness scenario, the central device, e.g. a smartphone, scans the frequencies for connectable advertising packets. The peripheral devices, e.g. the heart-rate sensor sends connectable advertising packets periodically and accepts incoming connections. Once a connection is established master and slave use generic attribute (GATT) profiles to exchange data. GATT is a simple structured list. Indeed, the data in GATT is organized in services and each service contains one or more characteristics.
Each characteristic consists of a universally unique identifier (UUID), a value and a set of properties. Bluetooth SIG defined UUIDs to identify BLE manufacturers as well~\cite{bluetooth-uuid}. By reading UUIDs data, the hacker might gather useful information to plan his attack.

\subsection{BLE Security}
In the BLE architecture (Figure~\ref{FIG:BLE-STACK}) SM is responsible for paring, integrity, authentication, and encryption~\cite{2017:NIST-800-121,2018:active-mitm-tal}. SM distributes security keys between peers and provides cryptographic functionalities. 

NIST~$800$-$121$-R2 details the security capabilities of Bluetooth and makes recommendations to effectively secure these devices. BLE security is different from Bluetooth BR/EDR. Since the introduction of the BLE $4.0$, the protocol supports a $128$-bit advanced encryption standard–counter with CBC-MAC (AES-CCM)~\cite{2017:NIST-800-121}. Although AES is considered one of the most secure forms of encryption, the key exchange protocol is exploitable. Indeed, during the \emph{pairing} process devices in BLE $4.0$ and $4.1$ versions exchange a temporary key (TK) and use it to create a short-term key (STK). These keys are used to encrypt the communication. In this case, an attacker can eavesdrop the keys and then decrypt the connection. This is not the case for BLE version $4.2$ and beyond due to the introduction of a long-term key (LTK) which uses the  elliptic curve Diffie Hellman (ECDH) key exchange, which is proved to be secure under this type of attack~\cite{2017:NIST-800-121, 2018:security_bluetooth_MIT}.

BLE $4.0$ and $4.1$ devices use the secure simple pairing  (SSP) model, in which devices based on their input/output (I/O) capabilities, choose one method from 
\begin{itemize}
	\item[$\bullet$] \textbf{Just Works}: TK is all zeros;
	\item[$\bullet$] \textbf{Passkey}: TK is a six-digit number combination inserted by the user;
	\item[$\bullet$] \textbf{Out-of-Band (OOB)}: TK is exchanged through a different medium.
\end{itemize}

The SM can protect the connection from MitM when the operating system (OS) selects Passkey or OOB as paring method. On the other hand, Just Works method does not provide any protection against MitM, that can be exploited by potential hackers.

\section{Cybersecurity in a WBAN Fitness Scenario}
\label{SEC:Fitness-scenario}
Despite the Bluetooth SIG released the new  Bluetooth $5.0$, there is still a huge number of devices in use that utilizes older versions of Bluetooth, such as version $4.1$ and $4.2$~\cite{bluetooth-market-2018}. Based on a recent estimation last year there were $4$ billion BLE enabled devices using version $4.0$ or $4.1$~\cite{2018:security_bluetooth_MIT}. Based on this information, the paper addresses security issues in BLE $4.1$ devices.
%
\begin{figure}[!t]
	\centering
	\includegraphics[width=0.75\textwidth]{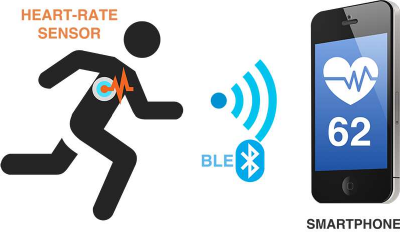}
	\caption{WBAN fitness scenario.} \label{FIG:FITNESS}
\end{figure}

As shown in Figure~\ref{FIG:FITNESS},  the scenario investigated in this paper includes a Polar H7 heart-rate BLE sensor worn at appropriate positions on the body~\cite{polar}. The device communicates a person's activity data through the BLE protocol to the smartphone. Due to their nature, WBAN might experience eavesdropping attacks. In this scenario, security and privacy are among major areas of concerns.

\begin{figure}[!b]
	\centering
	\includegraphics[width=0.75\textwidth]{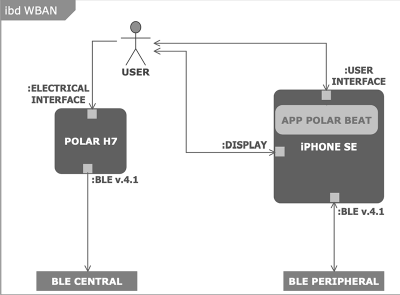}
	\caption{WBAN fitness interfaces representation with SysML.} \label{FIG:WBAN_SYSML}
\end{figure}

Once the author described the BLE protocol in the previous sections, the system analysis must be completed by describing the interfaces present in each device. Figure~\ref{FIG:WBAN_SYSML} shows the interfaces between the person, the Polar heart-rate sensor and the smartphone which runs the app to perform the synchronization. In particular, by using a SysML internal block diagram (IBD)~\cite{sysml}, the author highlighted only those interfaces that might have a key role in this analysis. 
%
%
\begin{table}[!t]
	\centering
	\begin{threeparttable}
		\renewcommand{\arraystretch}{1.3}
		\caption{Pairing procedure and I/O capabilities in the BLE fitness scenario}
		\label{TAB:PAIRING-FITNESS}
		\begin{tabular}{cccc}
			\multicolumn{1}{l}{} & \multicolumn{3}{c}{\textbf{Initiator}\tnote{1}} \\ \hhline{~-||--}
			\multicolumn{1}{c|}{\multirow{2}{*}{\begin{turn}{90}\textbf{Responder}\tnote{2}\end{turn}\hspace{5pt}}} & \multicolumn{1}{c||}{\textbf{\begin{tabular}[c]{@{}c@{}}I/O \\ \hspace*{5pt} Capabilities \hspace*{5pt}\end{tabular}}} & \multicolumn{1}{c|}{\textbf{\begin{tabular}[c]{@{}c@{}}Display \\ Only\end{tabular}}}  & \multicolumn{1}{c|}{\textbf{\begin{tabular}[c]{@{}c@{}}Keyboard \\ Display\end{tabular}}}  \\ \hhline{~=::==}
			\multicolumn{1}{c|}{} & \multicolumn{1}{c||}{\textbf{\begin{tabular}[c]{@{}c@{}}No Input \\  No Output\end{tabular}}} & \multicolumn{1}{c|}{\begin{tabular}[c]{@{}c@{}}Just Works\\ \hspace*{5pt} (Unauthenticated) \hspace*{5pt}\end{tabular} } & \multicolumn{1}{c|}{\begin{tabular}[c]{@{}c@{}}Just Works\\ \hspace*{5pt}(Unauthenticated) \hspace*{5pt}\end{tabular}} \\ \hhline{~|-||-|-|} 
		\end{tabular}
		\begin{tablenotes}
			\item [1] Smartphone;
			\item [2] Heart-rate BLE sensor;
		\end{tablenotes}
	\end{threeparttable}
\end{table}

The author has already discussed herein how the pairing procedure in BLE $4.1$ makes these devices prone to eavesdropping attacks.
The use of each association model is based on the I/O capabilities of the device. Considering that the smartphone, i.e. the \emph{initiator}, is equipped with a display and a keyboard display that can be used for pairing by the user. Whereas, the heart-rate sensor, i.e. the \emph{responder}, does not have any I/O capability.  Thus, in the scenario under investigation and by analyzing the available interfaces, Table~\ref{TAB:PAIRING-FITNESS} represents all possible pairing options. Accordingly, with the SSP model in this scenario, only the Just~Works method can be used. The user is required to accept a connection without verifying TK value on both devices, so Just~Works provides no MitM protection because it is an \emph{unauthenticated} pairing.

The common type of attacks against BLE communications are
\begin{itemize}
	\item[$\bullet$] \textbf{MitM} in which an attacker has the ability to both monitor and alter or inject messages into a communication channel;
	\item[$\bullet$] \textbf{Passive Eavesdropping} in which  the attacker is secretly listening (by using a sniffing device) to the private communication of others without consent.
\end{itemize}

\subsection{BLE $4.1$ Assessment}
\label{SEC:ASS}
This section describes the \emph{cybersecurity assessment} of BLE $4.1$ devices in a fitness scenario as shown in Figure~\ref{FIG:FITNESS}. The assessment combined multiple methodologies to best fit the investigation needs.
%
\begin{table}[!b]
	\centering
	\begin{threeparttable}
		\renewcommand{\arraystretch}{1.2}
		\caption{BLE $4.1$ vulnerability n.1}
		\label{TAB:VULN-1}
		\begin{tabularx}{\textwidth}{|b|m|}
			\hline
			\multicolumn{2}{|c|}{\textbf{Vulnerability n.1}} \\ \hline
			\textbf{Vulnerability} & \textbf{Low energy legacy pairing provides no passive eavesdropping protection.} \\ \hline
			\textbf{Likelihood} & High \\ \hline
			\textbf{Technical Impact} &  High\\ \hline
			\textbf{Risk} &  Critical\\ \hline
			\textbf{Threat Event} & Passive Eavesdropping \\ \hline
			\textbf{Description} & Eavesdroppers can capture secret keys (i.e., LTK) distributed during low energy pairing. \\ \hline
			\textbf{Mitigation} & BLE devices should be paired by using an algorithm that  provides a mechanism to exchange keys over an unsecured channel. For instance the ECDH.
			 \\ \hline
		\end{tabularx}
	\end{threeparttable}
\end{table}
The author selected NIST~$800$ series to evaluate threats and vulnerabilities of BLE sensors~\cite{2017:NIST-800-121, 2012:NIST-800-30}. Furthermore, the OWASP guideline is adopted to rate the risk associated with each vulnerability~\cite{2014:owasp_testing_guide}.

\Cref{TAB:VULN-1,TAB:VULN-2,TAB:VULN-3,TAB:VULN-4,TAB:VULN-5} report found vulnerabilities for the scenario under investigation. By following the methodology selected each table ties together concepts such as likelihood, technical impact, risk and threat events that could exploit the vulnerability. Moreover, for each vulnerability, the author provides a description and possible mitigation. 
Since these vulnerabilities have a risk mainly rated between \emph{high} and \emph{critical}, it should raise some concerns about the security of the heart-rate information transmitted by the sensor.
%
\begin{table}[!t]
	\centering
	\begin{threeparttable}
		\renewcommand{\arraystretch}{1.2}
		\caption{BLE $4.1$ vulnerability n.2}
		\label{TAB:VULN-2}
		\begin{tabularx}{\textwidth}{|b|m|}
			\hline
			\multicolumn{2}{|c|}{\textbf{Vulnerability n.2}} \\ \hline
			\textbf{Vulnerability} & \textbf{The Just Works pairing method provides no MITM protection.} \\ \hline
			\textbf{Likelihood} &  High\\ \hline
			\textbf{Technical Impact} & High  \\ \hline
			\textbf{Risk} & Critical \\ \hline
			\textbf{Threat Event} & MitM attack\\ \hline
			\textbf{Description} & MITM attackers can capture and manipulate data transmitted between trusted devices. \\ \hline
			\textbf{Mitigation} & Low energy devices should be paired in a secure environment to minimize the risk of eavesdropping and MITM attacks. Just Works pairing should not be used for low energy. \\ \hline
		\end{tabularx}
	\end{threeparttable}
\end{table}
%
\begin{table}[!t]
	\centering
	\begin{threeparttable}
		\renewcommand{\arraystretch}{1.2}
		\caption{BLE $4.1$ vulnerability n.3}
		\label{TAB:VULN-3}
		\begin{tabularx}{\textwidth}{|b|m|}
			\hline
			\multicolumn{2}{|c|}{\textbf{Vulnerability n.3}} \\ \hline
			\textbf{Vulnerability} & \textbf{No user authentication exists.} \\ \hline
			\textbf{Likelihood} & Medium \\ \hline
			\textbf{Technical Impact} & High \\ \hline
			\textbf{Risk} &  High\\ \hline
			\textbf{Threat Event} & Pairing Eavesdropping \\ \hline
			\textbf{Description} & Only device authentication is provided by the specification.  \\ \hline
			\textbf{Mitigation} & Application-level security, including user authentication, can be added via overlay by the application developer.\\ \hline
		\end{tabularx}
	\end{threeparttable}
\end{table}
%
\begin{table}[!h]
	\centering
	\begin{threeparttable}
		\renewcommand{\arraystretch}{1.2}
		\caption{BLE $4.1$ vulnerability n.4}
		\label{TAB:VULN-4}
		\begin{tabularx}{\textwidth}{|b|m|}
			\hline
			\multicolumn{2}{|c|}{\textbf{Vulnerability n.4}} \\ \hline
			\textbf{Vulnerability} & \textbf{End-to-end security is not performed.} \\ \hline
			\textbf{Likelihood} &  Medium\\ \hline
			\textbf{Technical Impact} & Medium \\ \hline
			\textbf{Risk} & Medium \\ \hline
			\textbf{Threat Event} & MitM attack \\ \hline
			\textbf{Description} & Only individual links are encrypted and authenticated. Data is decrypted at intermediate points. \\ \hline
			\textbf{Mitigation} & End-to-end security on top of the Bluetooth stack can be provided by use of additional security controls.\\ \hline
		\end{tabularx}
	\end{threeparttable}
\end{table}
%
\begin{table}[!h]
	\centering
	\begin{threeparttable}
		\renewcommand{\arraystretch}{1.2}
		\caption{BLE $4.1$ vulnerability n.5}
		\label{TAB:VULN-5}
		\begin{tabularx}{\textwidth}{|b|m|}
			\hline
			\multicolumn{2}{|c|}{\textbf{Vulnerability n.5}} \\ \hline
			\textbf{Vulnerability} & \textbf{Discoverable and/or connectable devices are prone to attack.} \\ \hline
			\textbf{Likelihood} & Medium \\ \hline
			\textbf{Technical Impact} &  High\\ \hline
			\textbf{Risk} &  High\\ \hline
			\textbf{Threat Event} & Passive Eavesdropping, MitM attack \\ \hline
			\textbf{Description} & A hacker can try to take over any  discoverable and/or connectable BLE device, and then he can get access to all the information. \\ \hline
			\textbf{Mitigation} & Any device that must go into discoverable or connectable mode to pair or connect should only do so for a minimal amount of time. A device should not be in discoverable or connectable mode all the time.\\ \hline
		\end{tabularx}
	\end{threeparttable}
\end{table}

\subsection{Experiment with MitM Attack}
Based on the cybersecurity assessment, in this section, the author proposes an active MitM attack for testing the BLE WBAN in a fitness scenario.

MitM usually involves three actors: Alice, Bob and Eve. In BLE networks this attacks changes its architecture. Indeed, the attacker, i.e. Eve, cannot act simultaneously as a sensor and as a mobile app.
Therefore, a BLE MitM needs to make use of two BLE components capable of acting together: one connects to the mobile app acting as the smartphone, while the other connects to the smartphone acting as the mobile app~\cite{2018:active-mitm-tal}. 

\begin{figure}[!t]
	\centering
	\includegraphics[width=0.75\textwidth]{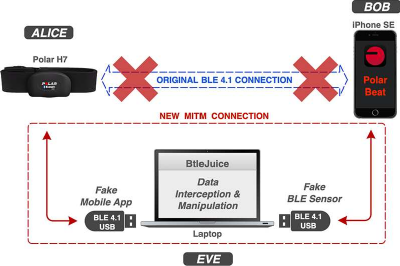}
	\caption{Active MitM architecture for BLE fitness scenario.} \label{FIG:MITM}
\end{figure}

The WBAN experiment setup consisted of a Polar H7 heart-rate sensor, i.e. Alice, and an Apple iPhone~SE~\cite{apple-iphone-se}, i.e. Bob.  The synchronization between the sensor and the smartphone is performed over BLE $4.1$. The smartphone ran the Polar Beat mobile app for real-time heart-rate monitoring~\cite{polar-beat}.
As shown in Figure~\ref{FIG:MITM}, Eve consists of a laptop that runs Linux Ubuntu $18.10$ and two CSR $8510$-based USB dongles that support BLE~$4.1$. These two dongles are connected to the laptop and communicate with each other using the \emph{BtleJuice} web-based software~\cite{btlejuice}. 

%
BtleJuice is a framework to perform MitM attacks on BLE devices. This framework consists of two parts which run on two virtual machines hosted by the same laptop. These parts named \emph{interception core} and \emph{proxy}, and they implement the MitM architecture shown in Figure~\ref{FIG:MITM}. And to do this, each virtual machine manages one USB dongle.

\begin{figure}[!b]
	\centering
	\includegraphics[width=0.75\textwidth]{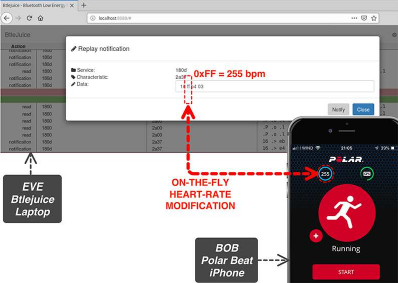}
	\caption{Heart-rate on-the-fly modification by using replay feature in Btlejuice.} \label{FIG:BTLEJUICE}
\end{figure}
BtleJuice acts as a proxy between the mobile app and the BLE heart-rate sensor. Any command sent to the sensor is captured by BtleJuice and relayed to the sensor. In particular, the interception proxy interacts with BLE peripherals and the interception core generates the fake devices with a fake BLE address. Then, the attacker from the web user interface (UI) can control the interception core. He can select the BLE target and intercept GATT operations. From the UI, it is possible to replay any GATT operation, but it allows also on-the-fly (OTF) data modification.

In WBAN BLE fitness scenario, Polar H7 heart-rate sensor communicates with the Polar Beat. Since the pairing process is completed, the author started  the experiment by attempting to actively sniff BLE traffic.
Figure~\ref{FIG:BTLEJUICE} shows how Eve, by using BtleJuice, can intercept the BLE information exchanged between the peripheral and the mobile app and manipulate data OTF. 

Once Btlejuice is initialized, the UI allows Eve the selection and the connection of the Polar H7 BLE sensor. In this way, Bob rather than connect to its peripheral, he connects his mobile app to the fake device. As shown in Figure~\ref{FIG:BTLEJUICE}, the attacker by using the replay function in Btlejuice can modify the heart-rate measurement, i.e. characteristics $0$x$2A37$~\cite{bluetooth-GATT-Char}, inside the heart-rate service, i.e. $0$x$180D$~\cite{bluetooth-GATT-Services}. As proof of concept (POC), Eve pushes $255$~beats per minute (bpm) in the Polar Beat app.

\subsection{Discussion about Security Countermeasures}
As pointed out during the cybersecurity assessment (Section~\ref{SEC:ASS}), BLE specifications do not offer defenses against MitM attacks. Although the experiment is limited to the WBAN fitness scenario described in this paper, these security leaks can give an advantage to the attacker.

On the L2CAP layer, there is the possibility to request an echo from the BLE sensor to measure round trip time (RTT) on the established link by using the \emph{l2ping} command. It is included in the BlueZ utils\cite{bluez}. The author assumes that the RTT of the MitM connection in Figure~\ref{FIG:MITM} is greater than the one in the BLE original connection. The evaluation of the RTT might offer a mechanism to detect the MitM. Unfortunately, the \emph{l2ping} command is not supported in most of the BLE peripherals. 

\begin{table}[!b]
    \centering
    \caption{RSSI measurements}
    \label{TAB:RSSI}
    \renewcommand{\arraystretch}{1.2}
    \begin{tabularx}{0.9\textwidth}{ | >{\centering\arraybackslash}X | >{\centering\arraybackslash}X | >{\centering\arraybackslash}X | >{\centering\arraybackslash}X | >{\centering\arraybackslash}X | }
        \hline
        \multirow{2}{*}{\textbf{RSSI}} & \multicolumn{4}{c|}{\textbf{Distance [m]}} \\ \cline{2-5} 
        & 0 & 0.5 & 1 & 3 \\ \hline
        \textbf{\begin{tabular}[c]{@{}c@{}}Mean\\ {[}dBm{]}\end{tabular}} & -26.4 & -52.8 & -60.8 & -66 \\ \hline
        \textbf{\begin{tabular}[c]{@{}c@{}}Std. Dev.\\ {[}dB{]}\end{tabular}} & 1.2 & 3.3 & 2.6 & 3 \\ \hline
    \end{tabularx}
\end{table}

Alternatively, the evaluation of the RSSI might offer another way to detect a MitM attack.
In the literature, there are several contributions to the relationship between the RSSI and the distance. In~\cite{2016:ble-rssi, 2017:ble-rssi-sensors-key}, RSSI is described as follows
%
%
\begin{equation}
\label{EQ:RSSI}
RSSI = - 10 \cdot N \cdot log(d) + a,
\end{equation} 
where $N$ is a constant assumed $1$, $d$ is the distance in meters and $a$ is the transmitted power at $1$-meter distance.

\begin{figure}[!t]
	\centering
	\includegraphics[width=0.75\textwidth]{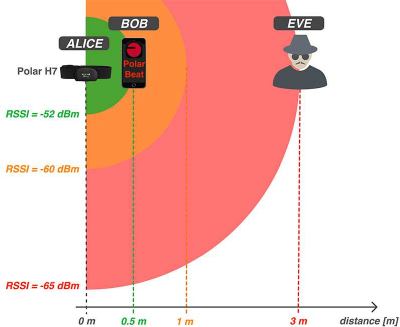}
	\caption{RSSI .} \label{FIG:RSSI}
\end{figure}
The author measured the RSSI by using the iPhone and the Bluefruit mobile app~\cite{bluefruit-app}. Table~\ref{TAB:RSSI} shows the average value and the standard deviation of the RSSI over $10$ measurements for each distance. These measurements validate the RSSI model in (\ref{EQ:RSSI}). Although the indoor environment and layout settings have a direct effect on the RSSI variability, the author assumes that Eve's RSSI can be greater than Bob's RSSI. Figure~\ref{FIG:RSSI} and Table~\ref{TAB:RSSI} confirm the assumption. In this scenario, by monitoring the RSSI value Bob can have a mechanism to detect the attack. Indeed, if the RSSI increased unexpectedly then the mobile app might alert the user about a possible attack.

\section{Conclusions}
\label{SEC:conclusions}
This paper aims to raise a concern about the need to be aware on the use of BLE devices. The author analyzed the security issues of BLE~$4.1.$ based sensors. Using the NIST classified threats, the author has identified a list of attacks which apply to BLE devices.

The WBAN scenario under test consists of a Polar H7 heart-rate sensor that communicates with the Polar Beat mobile app using the BLE technology. Being Polar Electro a top player in the wearable smart health devices, its BLE-based sensors are spread worldwide. With this research, the author would raise awareness about the security of the heart-rate information that we can receive from our wireless body sensors.

The case study shows that an attacker can easily intercept and manipulate the data transmitted between the mobile app and the BLE device.  Btlejuice was used to implement an active MitM attack, an operation that can result in the OTF modification of the data. The author remarks the importance to detect this kind of attack that might modify sensitive information such as the heart-rate.

%
%
\bibliographystyle{splncs04}
\bibliography{Bibtek.bib}
\end{document}